\documentclass[conference]{IEEEtran}
\IEEEoverridecommandlockouts
\usepackage{cite}
\usepackage{amsmath,amssymb,amsfonts}
\usepackage{algorithmicx}
\usepackage{graphicx}
\usepackage{textcomp}
\usepackage{xcolor}
\usepackage{algorithm}
\usepackage{algpseudocode}
\usepackage{mathtools}
\usepackage{subcaption}
\usepackage{xcolor}

\def\BibTeX{{\rm B\kern-.05em{\sc i\kern-.025em b}\kern-.08em
    T\kern-.1667em\lower.7ex\hbox{E}\kern-.125emX}}

\begin{document}

\title{  {Mapped ADMM: A Robust Algorithm for 1-Bit mMIMO Detection} \\
}

\author{Mohammad~Amin~Keshmiri,~
        Masoud~Ardakani,~\IEEEmembership{Senior~Member,~IEEE},
        
        \thanks{M.A. Keshmiri and M. Ardakani are with the Department of Electrical and Computer Engineering, University of Alberta, Edmonton, AB T6G 2R3, Canada (e-mails: keshmir1@ualberta.ca; ardakani@ualberta.ca).}

}

\maketitle

\begin{abstract}

Recently, it has been reported that one-bit massive MIMO (mMIMO) detection is equivalent to a binary classification problem that can be solved efficiently using support vector machine (SVM). Inspired by this result, we first reformulate SVM in a decentralized form consisting of multiple classifiers. This enables the use of the consensus alternating direction method of multipliers (CADMM), a technique that can improve robustness and performance through its inherent consensus making. We further update CADMM to output only valid constellation points and achieve significantly improved detection performance. In our method, by changing the size of the grouped classifiers, we balance the number of classifiers for consensus accuracy with sufficient data per group to ensure classifier robustness. Ultimately, we demonstrate that our proposed method significantly outperforms existing practically feasible methods for one-bit mMIMO detection.
\end{abstract}

\begin{IEEEkeywords}
Massive MIMO, One-bit detection, Binary classification, Support vector machine, Decentralization.  

\end{IEEEkeywords}

\section{Introduction}

Massive multi-input multi-output (mMIMO) technology has been one of the major breakthroughs of the last decade. It enables higher data rates and improved reliability by using many more receiver antennas than transmitter antennas, often nearly 10 times more \cite{7112627}. Due to its high spectral and energy efficiency, mMIMO has become a key technology for emerging systems such as terahertz and cell-free communications, particularly in the context of 6G \cite{10071970},\cite{11049871}.

Recently, interest has grown in scaling the number of antennas beyond conventional mMIMO, leading to ultra mMIMO (uMIMO) systems with thousands or even tens of thousands of antennas \cite{article}. Although increasing the number of antennas improves data rates, implementation cost and energy consumption remain major challenges. In particular, each antenna requires an analog-to-digital converter (ADC) at the receiver, and the power consumption and cost of high-resolution ADCs increase significantly with the number of quantization bits \cite{electronics13204128}.

A practical solution for these systems is to employ low-resolution ADCs, especially one-bit ADCs, which allow radical complexity reduction \cite{7600443}. However, one-bit quantization makes accurate data detection with conventional linear detectors almost impossible. As a result, many studies have focused on developing one-bit detectors that maintain satisfactory performance with manageable complexity.

Numerous methods have been proposed for one-bit data detection for mMIMO systems. Maximum likelihood (ML) detection, first discussed in \cite{7439790}, is optimal in performance but impractical because of its exponential exhaustive search complexity. Approximate message passing (AMP) and its variants, such as GAMP, variational AMP, and bi-linear GAMP, have been proposed to bypass the exponential complexity while maintaining near-optimal performance \cite{7094595},\cite{10599337}. However, these methods still suffer from high complexity \cite{9187842},\cite{10.5555/AAI28863751}.

Another group of methods formulates one-bit detection as an optimization problem based on the ML objective and solves a convex relaxation using gradient descent \cite{7439790}. Since gradient descent alone is generally insufficient, these methods usually require a second stage, such as local search over a reduced candidate space \cite{7439790} or an error recovery and correction process \cite{10843408}. Moreover, gradient-descent-based ML solvers suffer from severe degradation at high signal-to-noise ratios (SNRs), where the gradient of the ML objective becomes very small and prevents reliable convergence \cite{9148630}.

A different low-complexity approach was proposed in \cite{9148630},\cite{9387589}, where one-bit detection is formulated using a support vector machine (SVM) instead of the ML objective. This was done through an insightful observation, linking binary classification to one-bit detection. This method avoids the high-SNR limitations of the relaxed ML formulation, but as we see later, relying solely on non-differentiable gradients can still hinder the performance.

In this paper, we propose a new one-bit detection method that avoids the high computational complexity of ML- and AMP-based approaches, as well as the high-SNR degradation of relaxed ML-based methods, while surpassing the accuracy of the standard SVM-based detector. We first reformulate the SVM in a decentralized form, where antennas, either individually or grouped together with a chosen group size, provide multiple candidate estimates of the transmitted message. These candidate estimates can be aggregated using the consensus alternating direction method of multipliers (CADMM), leading to a more robust final detection, reducing the impact of noisy or erroneous local estimates. Furthermore, we accelerate the proposed method by using the knowledge that the transmitted symbols belong to a predefined constellation. Our constellation-aware modified CADMM leverages classifier group size as a hyperparameter to fine-tune the performance and complexity of our proposed mMIMO detection algorithm. Simulation results show that our proposed method completely overcomes the high-SNR degradation inherent in relaxed ML formulations and significantly outperforms existing SVM-based detectors at a lower computational complexity.

\section{System Model and Problem Definition}

Here, we first present the system model and then the SVM formulation for an uplink one-bit system.

\subsection{One-bit System Model}
We consider an uplink mMIMO system with $K$ users and $N_r$ antennas at the base station (BS). The channel state information is assumed to be perfectly known. Thus, we have
\begin{equation}
    \mathbf{r} = \mathbf{H}\mathbf{x} + \mathbf{n},    \label{eq1}
\end{equation}
where \(\mathbf{r} = [r_1, r_2, \dots, r_{N_r}]\) is the received vector, with each element corresponding to the reception at one base station antenna, and \(\mathbf{x} = [x_1, x_2, \dots, x_K]\) is the transmitted vector, with $x_j \in \mathcal{S}$ representing the transmitted symbol by user $j,1\le j\le K$ where $\cal S$ is the constellation from which the transmitted symbols are drawn. {\(\mathbf{H}^{N_r\times K}\) denotes the channel matrix, whose elements \(h_{ij}\), the channel coefficient between antenna $i$ and user $j$, follow a Rayleigh distribution and  \(\mathbf{n} = [n_1, n_2, \dots, n_{N_r}]\) is the noise vector, where $n_i$ represents the Gaussian noise at antenna} $i$.

For ease of analysis, the system model is transformed from the complex domain to the real domain. By placing the real part of the received signal, followed by the imaginary part into a vector, \eqref{eq1} is reformulated as:
\begin{equation}
    \mathbf{r}^\Re = (\mathbf{H}^\Re)^T\mathbf{x^\Re} + \mathbf{n}^\Re,  \label{eq2}
\end{equation}
where  $\mathbf{r}^\Re =[\Re\{\mathbf{r}\}^T,\Im\{\mathbf{r}\}^T]^T$, $\mathbf{x}^\Re =[\Re\{\mathbf{x}\}^T,\Im\{\mathbf{x}\}^T]^T,$  $\mathbf{n}^\Re =[\Re\{\mathbf{n}\}^T,\Im\{\mathbf{n}\}^T]^T,$ $\mathbf{H}^\Re =\begin{bmatrix}
    \Re\{\mathbf{H}\} & \Im\{\mathbf{H}\} \\
    -\Im\{\mathbf{H}\} & \Re\{\mathbf{H}\}
\end{bmatrix},$ and we have $\mathbf{H}^\Re =[\mathbf{h}_1,\mathbf{h}_2,...,\mathbf{h}_{2N_r}]. $

Since the received data is quantized to one bit, the sign of every reception is considered for data detection. That is

\begin{equation}
    \mathbf{y} = sgn(\mathbf{r}^\Re) = \begin{cases}1\hspace{4mm} \mathbf{r}^\Re \ge 0\\-1\hspace{4mm} \mathbf{r}^\Re<0\end{cases}.
    \label{eq3}
\end{equation}

\subsection{SVM-based detection Approach}

We now review the relation between binary classification and one-bit detection to establish a foundation for our proposed solution in Section III.

\subsubsection{Binary Classification and Data Detection}

In the considered system, the real-valued transmitted vector, $\mathbf{x}^{\Re}$, is observed through $2N_r$ virtual channels. After quantization, the received signal on the $i$-th path is
\begin{equation}
y_i=\operatorname{sgn}\left((\mathbf{x}^{\Re})^T \mathbf{h}_i + n_i\right).
\label{eq4}
\end{equation}
Given $\mathbf{h}_i$ and $y_i$, the detection goal is to recover $\mathbf{x}^{\Re}$.

To better see the connection between detection and binary classification, consider the noise-free case:
\begin{equation}
y_i=\operatorname{sgn}\left((\mathbf{x}^{\Re})^T \mathbf{h}_i\right).
\label{eq5}
\end{equation}
In this case, the channel vectors $\mathbf{h}_i$ are divided into two classes according to the sign of $y_i$, and these two classes are linearly separable by the hyperplane $(\mathbf{x}^{\Re})^T \mathbf{h}_i=0$. Therefore, one-bit detection can be interpreted as a binary classification problem in which $\mathbf{h}_i$ are the input samples, $y_i$ are the class labels, and $\mathbf{x}^{\Re}$ is the classifier parameter to be estimated.

\subsubsection{SVM-aided Formulation for Data Detection}

Since one-bit detection can be interpreted as a binary classification problem, prior work \cite{9387589} formulated the detection task as a soft-margin SVM problem, where the real-valued transmit vector $\mathbf{x}^{\Re}$ is treated as the classifier coefficient with zero bias:
\begin{equation}
\begin{cases}
\min_{\mathbf{x}^{\Re},\xi_i} \|\mathbf{x}^{\Re}\|^2 + C\sum_{i=1}^{2N_r}\xi_i \\
\text{s.t. } y_i(\mathbf{h}_i^T\mathbf{x}^{\Re}) \geq 1-\xi_i,\quad \xi_i \geq 0.
\end{cases}
\label{eq6}
\end{equation}
We can write the above problem in hinge-loss form as:
\begin{equation}
\min_{\mathbf{x}^{\Re}} \; \|\mathbf{x}^{\Re}\|^2
+ C\sum_{i=1}^{2N_r} \max\{0,\,1-y_i(\mathbf{h}_i^T\mathbf{x}^{\Re})\}.
\label{eq7}
\end{equation}
This reformulation explicitly reveals the nonsmooth hinge-loss term, which can lead to performance issues in gradient-based optimization, as demonstrated in the simulations.

\section{Proposed Solutions}
In this section, we propose a new solution to \eqref{eq7} that offers superior performance than {existing gradient-based approaches.

\subsection{Decentralized Data Detection}
A key idea, in our approach, is to reformulate \eqref{eq7} in a decentralized form such that it generates multiple candidate classifiers, or equivalently, multiple estimates of the transmitted vector. We then aggregate these estimates to form a final detection. This aggregation is expected to improve decision robustness. To this end, the objective function in \eqref{eq7} is decomposed into a sum of local objective functions, each associated with one received sample. We have
\begin{equation}
\begin{aligned}
f(\mathbf{x}^{\Re})
&= \|\mathbf{x}^{\Re}\|^2 + C\sum_{i=1}^{2N_r} \max\{0,\,1-y_i(\mathbf{h}_i^T\mathbf{x}^{\Re})\} \\
&= \sum_{i=1}^{2N_r} \left( \frac{1}{2N_r}\|\mathbf{x}^{\Re}\|^2 + C\max\{0,\,1-y_i(\mathbf{h}_i^T\mathbf{x}^{\Re})\} \right) \\
&= \sum_{i=1}^{2N_r} f_i(\mathbf{x}^{\Re}),
\end{aligned}
\label{eq8}
\end{equation}
where
$f_i(\mathbf{x}^{\Re}) = \frac{1}{2N_r}\|\mathbf{x}^{\Re}\|^2 + C\max\{0,\,1-y_i(\mathbf{h}_i^T\mathbf{x}^{\Re})\}.
$

We can now consider \(2N_r\) subproblems that can be solved, each creating an estimate of the transmitted vector, denoted by \(\mathbf{x}_i^{\Re}\). These estimates are then aggregated into one final decision. We can also bundle the terms in \eqref{eq8} into groups of size $M$, where $M$ observations collectively create one estimate \(\mathbf{x}_i^{\Re}\)  for \(\mathbf{x}^{\Re}\). We then have $2N_r/M$ such estimates over which a consensus has to be enforced. While a larger group size improves estimation robustness, the reduction in the number of groups weakens consensus accuracy, necessitating an optimal trade-off point. In summary, we rewrite \eqref{eq8} as
\begin{equation}
\begin{aligned}
f(\mathbf{x}^{\Re})
&= \sum_{i=1}^{2N_r} f_i(\mathbf{x}^{\Re}) = \sum_{i=1}^{2N_r/M}\sum_{m=1}^M f_{\{M(i-1) +m\}}(\mathbf{x}^{\Re})\\
&=\sum_{i=1}^{2N_r/M}d_i(\mathbf{x}^{\Re})=\sum_{i=1}^{2N_r/M}d_i(\mathbf{x}_i^{\Re}),\quad (\mathbf{x}_i^{\Re} = \mathbf{x}^{\Re} \hspace{1mm}\forall i)\\
\end{aligned}
\label{eq9}
\end{equation}
where $d_i(\mathbf{x}_i^{\Re})=\sum_{m=1}^M f_{\{M(i-1) +m\}}(\mathbf{x}_i^{\Re}).$
Consequently, \eqref{eq9} can be reformulated as
\begin{equation}
\begin{cases}
\displaystyle \min_{\{\mathbf{x}_i^{\Re}\},\,\mathbf{x}^{\Re}} \sum_{i=1}^{2N_r/M}d_i(\mathbf{x}_i^{\Re}), \\
\text{s.t. } \mathbf{x}_i^{\Re} = \mathbf{x}^{\Re}, \quad \forall i.
\end{cases}
\label{eq10}
\end{equation}

A well-suited approach for solving \eqref{eq10} is CADMM that solves separable optimization problems with shared global variable. In this framework, instead of the conventional Lagrangian, augmented Lagrangian is utilized, written as
\begin{equation}
\begin{aligned}
&L_{\rho}\!\left(\{\mathbf{x}_i^{\Re}\},\mathbf{x}^{\Re},\{\boldsymbol{\lambda}_i\}\right)\\
&= 
\sum_{i=1}^{2N_r/M}d_i(\mathbf{x}_i^{\Re})
+\boldsymbol{\lambda}_i^{T}(\mathbf{x}_i^{\Re}-\mathbf{x}^{\Re}) 
+\frac{\rho}{2}\|\mathbf{x}_i^{\Re}-\mathbf{x}^{\Re}\|^2
,
\end{aligned}
\label{eq11}
\end{equation}
where \(\boldsymbol{\lambda}_i\) is the dual variable associated with the \(i\)-th consensus constraint, and \(\rho>0\) is the penalty parameter. From \eqref{eq11}, the CADMM update rule for the $i$-th subproblem is 
\begin{equation}
\begin{aligned}
&\mathbf{x}_i^{\Re}(t+1) = \\ 
&\arg\min_{\mathbf{x}_i^{\Re}} \Big( d_i(\mathbf{x}_i^{\Re}) + \boldsymbol{\lambda}_i^{T}(t)\bigl(\mathbf{x}_i^{\Re}-\mathbf{x}^{\Re}(t)\bigr)
 + \frac{\rho}{2}\|\mathbf{x}_i^{\Re}-\mathbf{x}^{\Re}(t)\|^2 \Big).
\end{aligned}
\label{eq12}
\end{equation}
This can be solved using
\begin{equation}
\mathbf{x}_i^{\Re}(t+1,k+1) = \mathbf{x}_i^{\Re}(t+1,k) - \alpha \,\mathbf{g}_i(t+1,k),
\label{eq13}
\end{equation}
where $\alpha$ is the learning rate and $\mathbf{g}_i(t+1,k)$ is the subgradient of the local augmented objective at the $k$-th iteration of an inner loop to update local variables. Also, the subgradient is 
\begin{equation}
\begin{aligned}
\mathbf{g}_i(t+1,k) = {} & \frac{1}{N_r}\mathbf{x}_i^{\Re}(t+1,k)\\
&- C\sum_{m=1}^M \mathbf{1}_{m,i} y_{M(i-1)+m}\mathbf{h}_{M(i-1)+m} \\
& + \boldsymbol{\lambda}_i(t) + \rho\bigl(\mathbf{x}_i^{\Re}(t+1,k)-\mathbf{x}^{\Re}(t)\bigr),
\end{aligned}
\label{eq14}
\end{equation}
where 
$\mathbf{1}_{m,i} = 
\begin{cases}
1, & \text{if } y_{M(i-1)+m}\mathbf{h}_{M(i-1)+m}^T\mathbf{x}_i^{\Re}(t+1,k) < 1, \\
0, & \text{otherwise},
\end{cases}$
and $\mathbf{x}_i^{\Re}(t+1,k)$ denotes the $k$-th iteration of the inner loop to compute $\mathbf{x}_i^{\Re}(t+1)$.

 Subgradient $\mathbf{g}_i(t+1,k)$ is non-smooth because the subproblem \eqref{eq12} is non-differentiable. To resolve this, our algorithm uses a consensus-building outer loop that enforces the constraint $\mathbf{x}_i^{\Re} = \mathbf{x}^{\Re}$ across all subproblems. The outer loop is:
\begin{equation}
\mathbf{x}^{\Re}(t+1)
=
\frac{1}{2N_r}
\sum_{i=1}^{2N_r}
\left(
\mathbf{x}_i^{\Re}(t+1)
+
\frac{1}{\rho}\boldsymbol{\lambda}_i(t)
\right),
\label{eq15}
\end{equation}
and the dual variables are then updated according to: \begin{equation}
\boldsymbol{\lambda}_i(t+1)
=
\boldsymbol{\lambda}_i(t)
+
\rho\bigl(\mathbf{x}_i^{\Re}(t+1)-\mathbf{x}^{\Re}(t+1)\bigr).
\label{eq16}
\end{equation}

This outer loop results in averaging across subproblems, reducing the impact of highly corrupted observations. This consensus mechanism improves robustness compared to a single global classifier and yields a more accurate estimate of the transmitted vector.

After convergence, each element of the transmitted vector is recovered by projecting the final estimate onto the nearest point in the signal constellation:
\begin{equation}
\hat{x}_j^{\Re}
=
\arg\min_{x\in\mathcal{S}}
\left|x_j^{\Re}-x\right|^2,
\label{eq17}
\end{equation}
where \(x_j^{\Re}\) denotes the \(j\)-th element of \(\mathbf{x}^{\Re}\), and $\hat{\mathbf{x}}^{\Re}$ denotes the corresponding projected vector whose $j$-th element is $\hat{x}_{j}^{\Re}$.

\textcolor{red}{
}

\subsection{Mapped ADMM}
We now modify our approach to accelerate our prior solution. Recall that CADMM seeks a real-valued estimate of $\mathbf{x}^{\Re}$. However, the elements of $\mathbf{x}^{\Re}$ belong to a finite constellation set. Naturally, we can exploit this inherent property. This, as seen later, will accelerate convergence and slightly improve performance. In particular, we suggest mapping each local estimate into a constellation point and then forcing the consensus based on a voting system. This is in contrast to our first method, in which, first a consensus is built and then the agreed-upon continuous variable is mapped to a constellation point. We call this method mapped ADMM (MADMM). Specifically, each subproblem's local variable, $\mathbf{x}_i^{\Re}(t+1)$, is mapped onto the constellation $\mathcal{S}$ symbol by symbol using the minimum-distance criterion:
\begin{equation}
    \hat{x}_{ij}^{\Re}(t+1) = \arg\min_{x \in \mathcal{S}} \left| x_{ij}^{\Re}(t+1) - x \right|^2,
    \label{eq18}
\end{equation}
where $x_{ij}^{\Re}(t+1)$ denotes the $j$-th element of $\mathbf{x}_i^{\Re}(t+1)$, and $\hat{\mathbf{x}}_{i}^{\Re}(t+1)$ denotes the corresponding mapped vector.

Subsequently, after performing this projection for all subproblems, a majority-voting scheme is employed to reach a consensus. The vector most frequently estimated across all subproblems is selected as the transmitted vector. To ensure decision reliability, the iterative process continues until the number of votes for the most frequent candidate exceeds that of the second-most frequent candidate by a predefined gap. This vote gap threshold, $\gamma$, establishes a trade-off between convergence speed and detection reliability. Even when the vote gap is set to its maximum possible value of $2N_r/M$, this strategy significantly accelerates the algorithm.
The MADMM method is summarized in Algorithm \ref{alg1}.

\begin{algorithm}
\caption{Pseudo Code of the Proposed MADMM}\label{alg1}

$\mathbf{Input} :\hspace{0.5mm} (\mathbf{h}_i,y_i)  \hspace{3mm} \forall i \in[1,2N_r]$

$\mathbf{Hyper Parameters} :\hspace{0.5mm} C, \rho, \alpha, \gamma,\epsilon$

$\mathbf{Output} :\hspace{0.5mm} {\mathbf{{x}}_{\text{MADMM}}}$

\begin{algorithmic}[1] 
    \State  $\gamma = 2N_r/M$
    
    \While{$\frac{\|\mathbf{x}^\Re{(t)} - \mathbf{x}^\Re{(t-1)}\|}{\|\mathbf{x}^\Re{(t-1)}\|} > \epsilon$}
        \For{$i = 1$ to ${2N_r}/{M}$}
            \While{$\frac{\|\mathbf{x}_i^\Re{(t,k)} - \mathbf{x}_i^\Re{(t,k-1)}\|}{\|\mathbf{x}^\Re_i{(t,k-1)}\|}>\epsilon$}

            \State $\mathbf{x}_i^{\Re}(t+1,k+1) =\mathbf{x}_i^{\Re}(t+1,k)-\alpha \,\mathbf{g}_i(t+1,k)$
            \State $ \mathbf{x}_i^{\Re}(t+1,k+1)\xleftarrow{}\frac{\sqrt{K}\mathbf{x}_i^{\Re}(t+1,k+1)}{\|\mathbf{x}_i^{\Re}(t+1,k+1)\|}$ 
            \State $ \mathbf{\hat{x}}_i^{\Re}(t+1,k+1)\xleftarrow{map}\mathbf{x}_i^{\Re}(t+1,k+1)$ 

            \State $ k\xleftarrow{}k+1$           
            \EndWhile
          \State Find the vectors $\mathbf{s}^*$ and $\mathbf{s}^{\dag}$ containing the most and second-most frequent constellation symbols in $\mathbf{\hat{x}}_i^{\Re}(t+1)$, with their respective frequencies denoted as $n_j^*$ and $n_j^\dag$.
             \State $\mathbf{x}^\Re{(t+1)} =\frac{1}{2N_r}\sum_{i=1}^{2N_r}(\mathbf{x}_i^{\Re}(t+1) +\mathbf{\lambda}_i(t))$
             \State $ \mathbf{x}^{\Re}(t+1)\xleftarrow{}\frac{\sqrt{K}\mathbf{x}^{\Re}(t+1)}{\|\mathbf{x}^{\Re}(t+1)\|}$ 
             
            \State\textbf{if} $\frac{1}{K}\sum_{j=1}^{K} (n_j^*-n_j^{\dag}) \geq\gamma$  \textbf{break;}

            \State $\mathbf{\lambda}_i(t+1)= \mathbf{\lambda}_i(t) + \rho(\mathbf{x}^\Re{(t+1)}-\mathbf{x}_i^\Re{(t+1)})$ $\forall i$
             \State $ t\xleftarrow{}t+1$           

        \EndFor
    \EndWhile
    
\State ${\mathbf{{x}}_{\text{MADMM}} \xleftarrow{map} \mathbf{s}^*}$
\end{algorithmic}

\end{algorithm}

\subsection{Complexity Analysis}
We will investigate the performance of MADMM and the impact of group size in the simulation result section. Here, we study the impact of group size on complexity. Empirically, increasing the group size reduces the number of subproblems to be solved by a factor of $M$. Within each subproblem, lines 5, 6, and 7 constitute the core computational operations. Due to the cumulative sum over $M$ required to generate the subgradient in \eqref{eq14}, the complexity of line 5 remains unchanged for all group sizes. However, for the other two operations, the number of required computations scales down by a factor of $M$. For simplicity of discussions, let us count each of lines 5, 6, and 7 as one operation. Then the total number of operations can be quantified as
\begin{equation}
 N_{\text{op}} = \frac{N_r}{M} \times (M +2) = N_r +2\frac{N_r}{M}.
\end{equation}
Therefore, the complexity grows linearly with the number of antennas, and increasing the group size reduces the computational overhead. However, increasing group size also reduces the total number of classifiers, meaning an erroneous classifier more heavily degrades the consensus performance. This will be seen in the numerical results. Hence, the choice of $M$ creates a trade-off between error performance and computational efficiency.

\section{Simulation Results}

In this section, we first compare our proposed MADMM and CADMM in terms of convergence speed and performance. Next, the impact of group size on performance and complexity is investigated. Finally, the performance of MADMM is evaluated against existing benchmarks. All simulations are performed using the Monte Carlo method with $ 10^5$ independent trials.

The convergence speeds of CADMM and MADMM are evaluated under two mMIMO configurations using QPSK modulation. For a $32 \times 4$ system ($N_r = 32$ BS antennas and $K = 4$ users), CADMM requires an average of 381 iterations to converge, whereas MADMM requires only 40. When scaling the system to a $64 \times 8$ configuration, these iteration counts shift to 618 for CADMM and 48 for MADMM, demonstrating that MADMM offers significantly better convergence scalability. Conversely, the standard SVM formulation fails to converge within the simulation framework, reaching the maximum limit of $10^5$ iterations. Furthermore, employing MADMM yields superior performance. By mapping directly into the constellation space, it solves the problem more accurately than real-space approaches, as illustrated across both mMIMO configurations in Fig. \ref{fig:comparison}.

\begin{figure}
    \centering
    \includegraphics[width=0.85\linewidth]{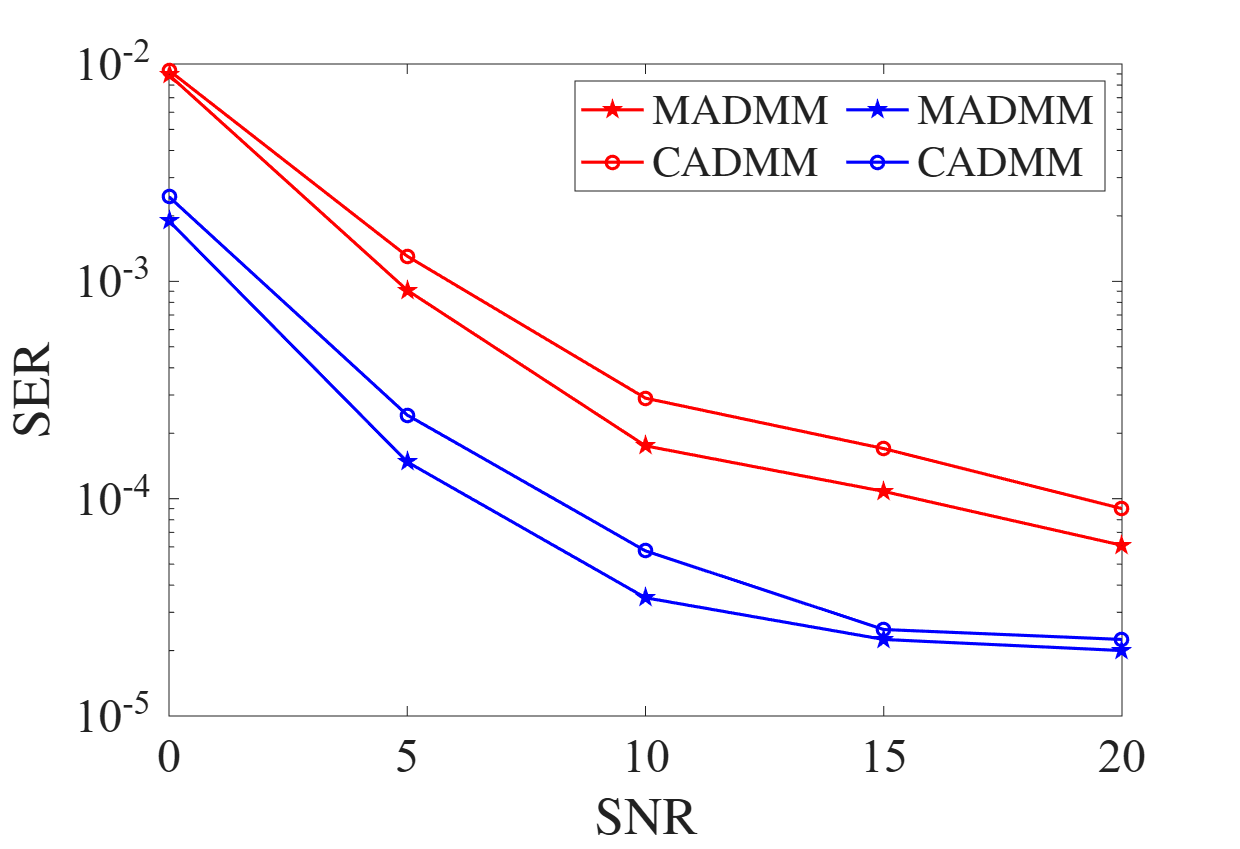}
    \caption{SER versus SNR performance of MADMM againts CADMM using QPSK modulation for $32\times4$ (red) and $64\times8$ (blue) mMIMO configurations.}
    \label{fig:comparison}
\end{figure}

MADMM can also navigate the complexity-performance trade-off via the group size. The previous section studied the impact of group size on complexity. Fig. \ref{fig:grouping} demonstrates the performance comparison for four different group sizes. While a smaller group size yields more local consensus candidates, it reduces the robustness of each individual estimate. Optimizing the group size mitigates this by balancing consensus accuracy with resilience against corrupted data. For instance, $M=4$ provides the best performance and lower complexity compared to $M=1$ and $M=2$ for the $64\times8$ configuration.

\begin{figure}
    \centering
    \includegraphics[width=0.85\linewidth]{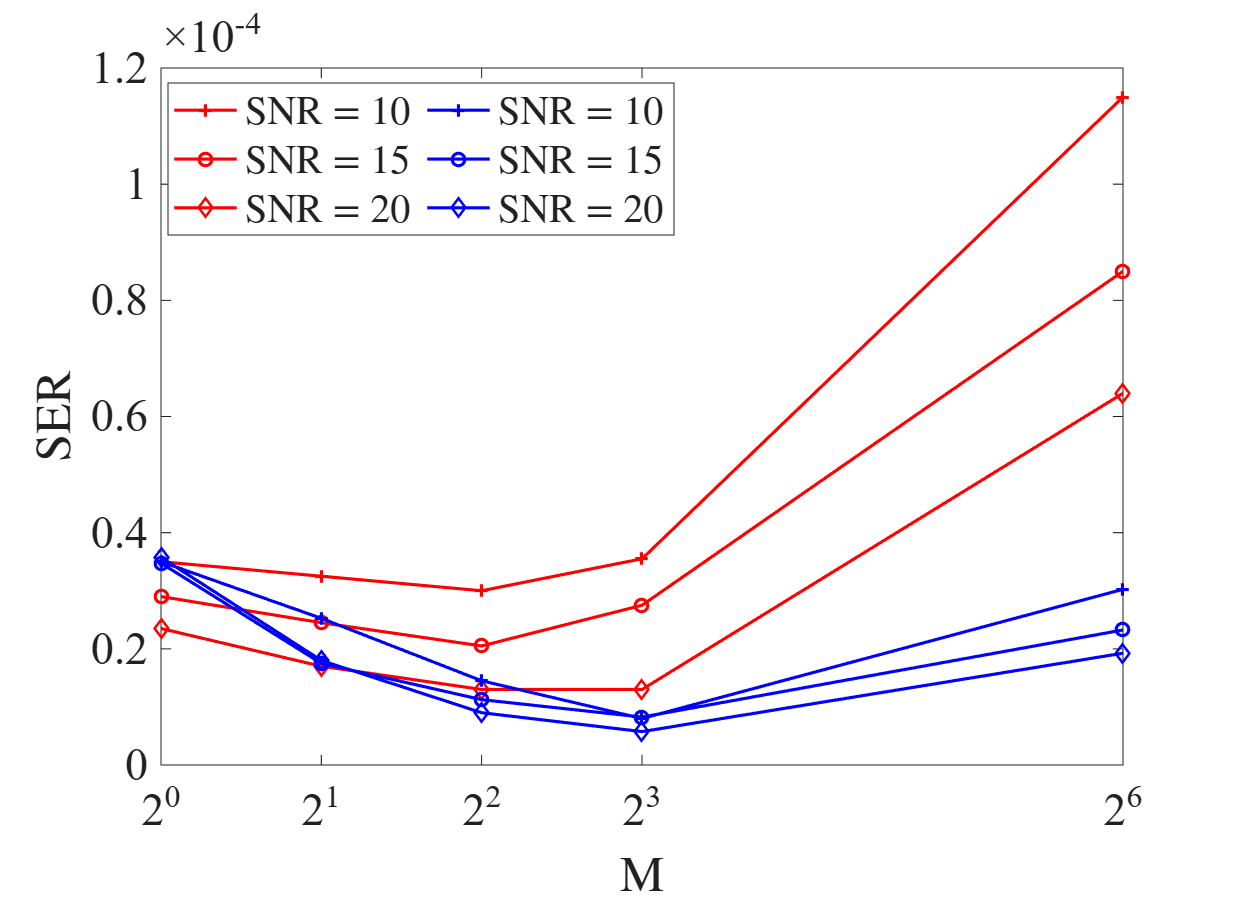}
    \caption{SER versus group size performance of MADMM across various SNRs using QPSK for $64\times8$ (red) and $128\times16$ (blue) mMIMO.}
    \label{fig:grouping}
\end{figure}

We now compare MADMM with two important one-bit detection benchmarks. The conventional zero-forcing (ZF) detector, which is the first one, is a linear scheme formulated as follows:
\begin{equation}
    \mathbf{x}_{\text{ZF}} = (\mathbf{H}^H\mathbf{H})^{-1}\mathbf{H}^H\mathbf{y}.
    \label{eq20}
\end{equation}
Subsequently, $\mathbf{x}_{\text{ZF}}$ is mapped to the discrete constellation set $\mathcal{S}$. 
The second benchmark \cite{7439790} bypasses intensive ML decoding by treating detection as a convex optimization problem, achieving high accuracy via gradient descent. As initially observed in \cite{9387589}, this method, known as near maximum likelihood (NML), suffers from performance degradation in the high-SNR regime. More specifically, the gradient of the ML objective function decays rapidly with SNR, thereby hindering reliable convergence.

 Figures \(\ref{fig:1}\) and \(\ref{fig:2}\) illustrate the SER performance of MADMM relative to the SVM, the NML and ZF detectors for \(32 \times 4\) and \(64 \times 8\) mMIMO configurations, respectively. As expected, ZF, as a linear detector, fails to estimate accurately compared with the proposed method. The results also highlight that MADMM outperforms SVM and maintains robust performance in the high-SNR regime, in contrast to NML, which suffers from significant performance degradation as the SNR increases.

\begin{figure}
    \centering
    \includegraphics[width=0.8\linewidth]{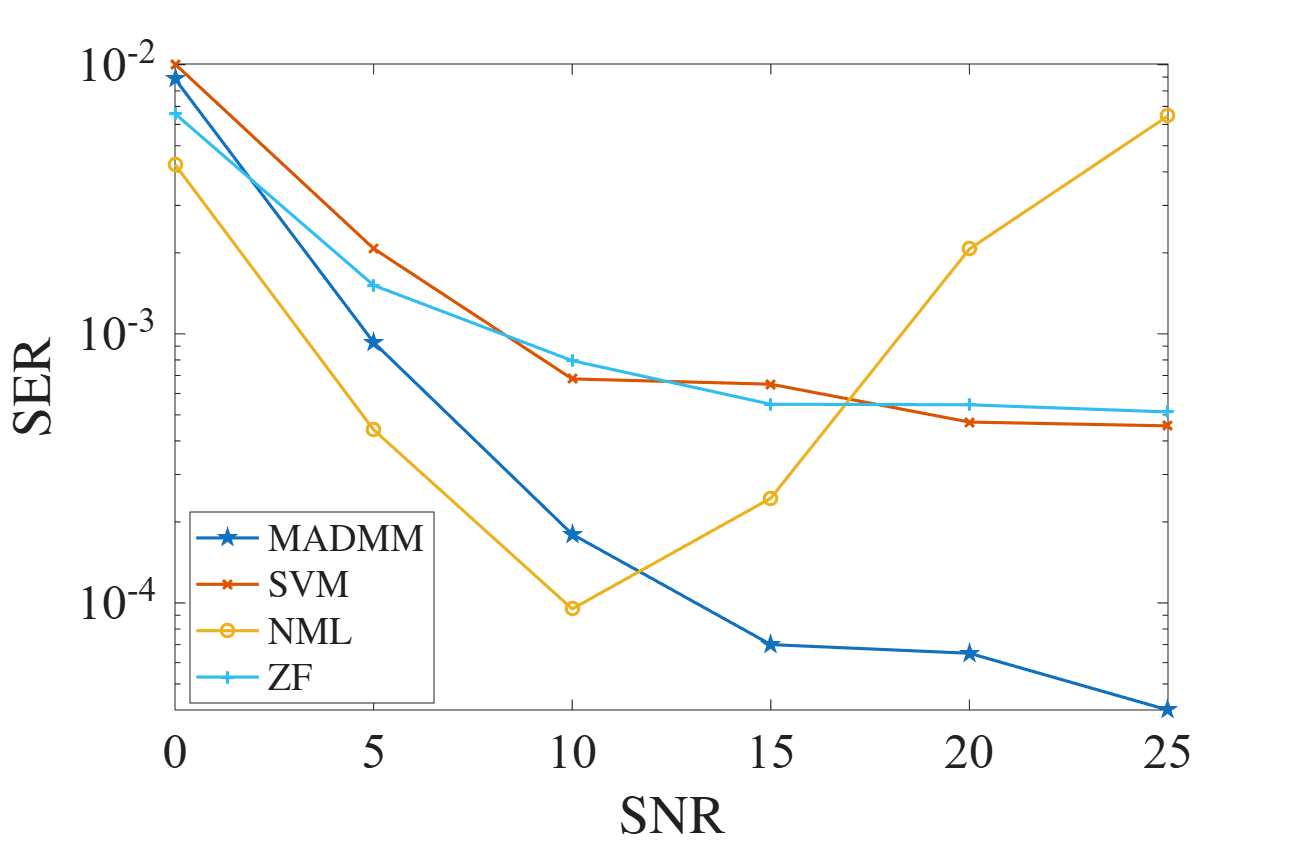}
    \caption{SER versus SNR performance of all one-bit detection methods using QPSK modulation for a $32\times4$ mMIMO.}
    \label{fig:1}
\end{figure}

\begin{figure}
    \centering
    \includegraphics[width=0.8\linewidth]{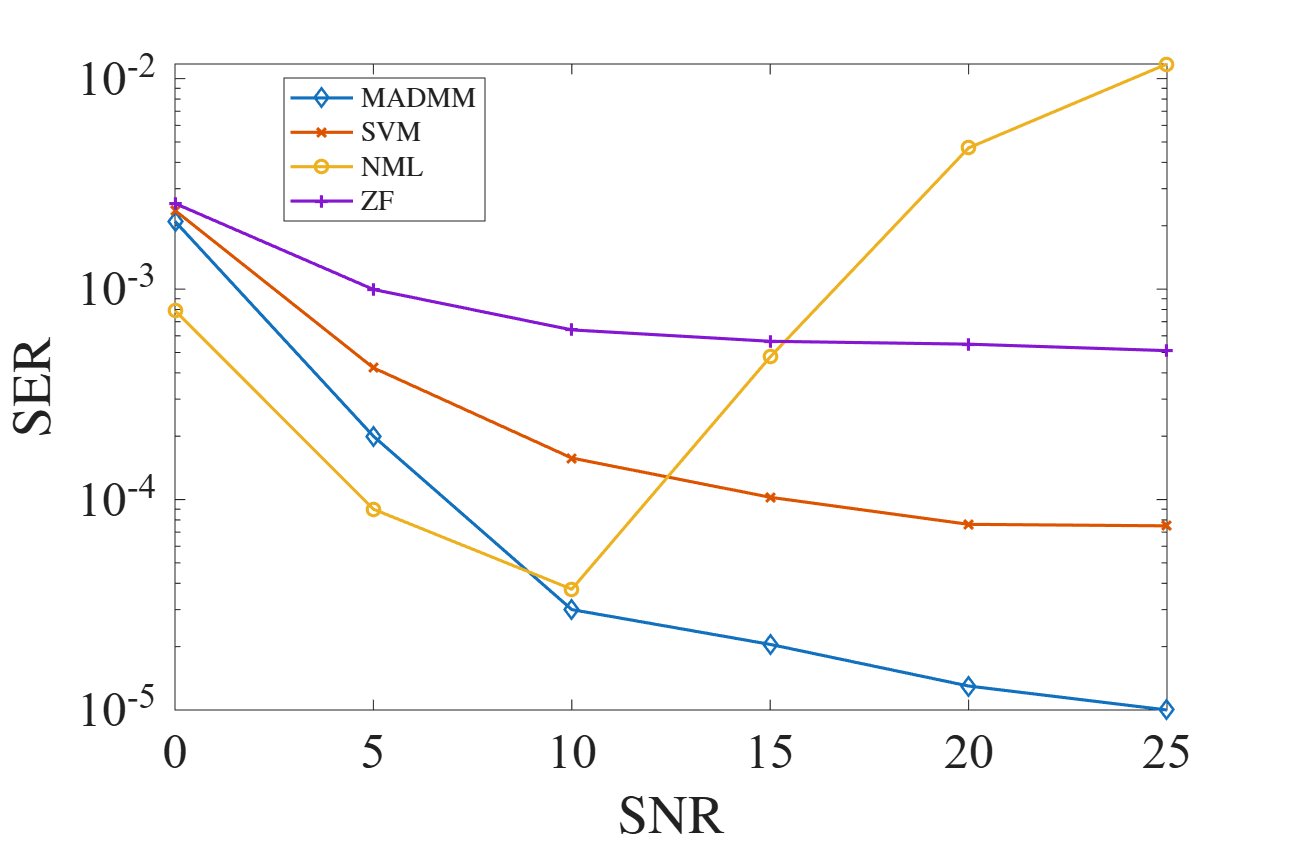}
    \caption{SER versus SNR performance of all one-bit detection methods using QPSK modulation for a $64\times8$ mMIMO.}
    \label{fig:2}
\end{figure}

\section{Conclusion}

In this paper, a novel detector is proposed for one-bit detection in mMIMO systems. The method is based on the similarity between one-bit detection and binary classification. The core idea is to reformulate the detection problem in a decentralized form, allowing for multiple detection candidates and enforcing consensus among these candidates. The decentralized form can be further grouped to adjust the number of detection candidates. Our decentralized approach improves the detection performance compared to existing methods and its grouping feature allows for navigating the performance-complexity landscape. We further improve the convergence speed of our proposed solution by proposing MADMM, which uses the inherent property that user signals belong to a constellation set. The idea of MADMM is to map the detected signal to the nearest constellation point before building consensus. MADMM significantly speeds up convergence compared to CADMM, our initial proposed solution.

\bibliographystyle{IEEEtran}
\bibliography{main.bib}

\end{document}